\documentclass[12pt]{article}

\begin{document}

\def\CA{{\cal A}}
\def\CB{{\cal B}}
\def\CC{{\cal C}}
\def\CD{{\cal D}}
\def\CE{{\cal E}}
\def\CF{{\cal F}}
\def\CG{{\cal G}}
\def\CH{{\cal H}}
\def\CI{{\cal I}}
\def\CJ{{\cal J}}
\def\CK{{\cal K}}
\def\CL{{\cal L}}
\def\CM{{\cal M}}
\def\CN{{\cal N}}
\def\CO{{\cal O}}
\def\CP{{\cal P}}
\def\CQ{{\cal Q}}
\def\CR{{\cal R}}
\def\CS{{\cal S}}
\def\CT{{\cal T}}
\def\CU{{\cal U}}
\def\CV{{\cal V}}
\def\CW{{\cal W}}
\def\CX{{\cal X}}
\def\CY{{\cal Y}}
\def\CZ{{\cal Z}}

\newcommand{\todo}[1]{{\em \small {#1}}\marginpar{$\Longleftarrow$}}
\newcommand{\labell}[1]{\label{#1}\qquad_{#1}} 
\newcommand{\bbibitem}[1]{\bibitem{#1}\marginpar{#1}}
\newcommand{\llabel}[1]{\label{#1}\marginpar{#1}}

\newcommand{\sphere}[0]{{\rm S}^3}
\newcommand{\su}[0]{{\rm SU(2)}}
\newcommand{\so}[0]{{\rm SO(4)}}
\newcommand{\bK}[0]{{\bf K}}
\newcommand{\bL}[0]{{\bf L}}
\newcommand{\bR}[0]{{\bf R}}
\newcommand{\tK}[0]{\tilde{K}}
\newcommand{\tL}[0]{\bar{L}}
\newcommand{\tR}[0]{\tilde{R}}

\newcommand{\btzm}[0]{BTZ$_{\rm M}$}
\newcommand{\ads}[1]{{\rm AdS}_{#1}}
\newcommand{\ds}[1]{{\rm dS}_{#1}}
\newcommand{\eds}[1]{{\rm EdS}_{#1}}
\newcommand{\sph}[1]{{\rm S}^{#1}}
\newcommand{\gn}[0]{G_N}
\newcommand{\SL}[0]{{\rm SL}(2,R)}
\newcommand{\cosm}[0]{R}
\newcommand{\hdim}[0]{\bar{h}}
\newcommand{\bw}[0]{\bar{w}}
\newcommand{\bz}[0]{\bar{z}}
\newcommand{\be}{\begin{equation}}
\newcommand{\ee}{\end{equation}}
\newcommand{\bea}{\begin{eqnarray}}
\newcommand{\eea}{\end{eqnarray}}
\newcommand{\pat}{\partial}
\newcommand{\lp}{\lambda_+}
\newcommand{\bx}{ {\bf x}}
\newcommand{\bk}{{\bf k}}
\newcommand{\bb}{{\bf b}}
\newcommand{\BB}{{\bf B}}
\newcommand{\tp}{\tilde{\phi}}
\hyphenation{Min-kow-ski}

\def\apr{\alpha'}
\def\str{{str}}
\def\lstr{\ell_\str}
\def\gstr{g_\str}
\def\Mstr{M_\str}
\def\lpl{\ell_{pl}}
\def\Mpl{M_{pl}}
\def\varep{\varepsilon}
\def\del{\nabla}
\def\grad{\nabla}
\def\tr{\hbox{tr}}
\def\perp{\bot}
\def\half{\frac{1}{2}}
\def\p{\partial}
\def\perp{\bot}
\def\eps{\epsilon}

\renewcommand{\thepage}{\arabic{page}}
\setcounter{page}{1}

\rightline{hep-th/0103171}
\rightline{CITUSC/01-005}
\rightline{RUNHETC-2001-09}
\rightline{UPR-929-T}
\vskip 1cm
\centerline{\Large \bf Deconstructing de Sitter}
\vskip 1cm

\renewcommand{\thefootnote}{\fnsymbol{footnote}}
\centerline{{\bf Vijay
Balasubramanian${}^{1,4}$\footnote{vijay@endive.hep.upenn.edu},
Petr Ho\v{r}ava${}^{2,4}$\footnote{horava@physics.rutgers.edu},
and
Djordje Minic${}^{3}$\footnote{minic@citusc.usc.edu}
}}
\vskip .5cm
\centerline{${}^1$\it David Rittenhouse Laboratories, University of
Pennsylvania}
\centerline{\it Philadelphia, PA 19104, U.S.A.}
\vskip .5cm
\centerline{${}^2$\it Department of Physics and Astronomy, Rutgers
University}
\centerline{\it Piscataway, NJ 08854-8019, U.S.A.}
\vskip .5cm
\centerline{${}^3$\it CIT-USC Center for Theoretical Physics}
\centerline{${}^3$\it Department of Physics and Astronomy, University
of Southern California}
\centerline{\it Los Angeles, CA 90089-0484, U.S.A.}
\vskip .5cm
\centerline{${}^4$\it Institute for Theoretical Physics, University of
California}
\centerline{\it Santa Barbara, CA 93106-4030, U.S.A.}

\setcounter{footnote}{0}
\renewcommand{\thefootnote}{\arabic{footnote}}

\begin{abstract}
Semiclassical gravity predicts that de Sitter space has a finite
entropy.  We suggest a picture for Euclidean de Sitter space in string
theory, and use the AdS/CFT correspondence to argue that  de Sitter entropy
can be understood as the number of degrees of freedom in a quantum
mechanical dual.
\end{abstract}


\section{Introduction}
\label{intro}

Current astronomical observations suggest that the universe is
well-approximated by a world with a positive cosmological constant
$\Lambda > 0$~\cite{expand}.  If this is the case, string theorists
must contend with two strange facts.  First, a small, positive
cosmological constant is surprisingly hard to construct within the
usual framework of string theory and the low energy supergravities
that arise from it.  Of course, we might expect that nothing prevents
the generation of a cosmological constant after supersymmetry is
broken.  However, finding a satisfactory SUSY breaking mechanism is
itself a difficult problem in string theory.  We might suppose that
this state of affairs is simply a failure of imagination among
theorists, but there is a more troubling fact.  De Sitter space is
expanding so rapidly that inertial observers see a cosmological
horizon, and application of the Bekenstein-Hawking bound to this
horizon implies that $D$ dimensional de Sitter space has an entropy $S
\sim \Lambda^{-(D -2)/2} / \gn$.  This is particularly disturbing
since it suggests that the entire observable universe should perhaps
have a finite number of degrees of freedom.  This interpretation would
certainly contradict both field theory and perturbative string theory.

A potential conclusion based on these facts, and particularly the
latter one, is that the world cannot harbour a positive cosmological
constant.  In that case, the astronomical data must be interpreted to
say that we are yet to settle into the true vacuum.  Either we
continue to roll slowly down a potential, realizing something like
quintessence (see, e.g.,~\cite{quint}), or else we await a sharp phase
transition in the future.

Regardless, during the epoch preceding descent to the true vacuum, the
world is rapidly expanding as though there is a cosmological constant,
and any inertial observer sees an apparent cosmological horizon.  In
recent years, evidence has arisen that the Bekenstein-Hawking entropy
of event horizons can be extended to a more local statement regarding
the entropy enclosed within, or encoded on, any closed
surface~\cite{holog,raphael1,emw}.  This suggests that during the
epoch of rapid expansion an inertial observer will appear to be in
contact with a universe of finite entropy.  We therefore return to the
question of whether a rapidly expanding space like de Sitter actually
has a finite entropy.  If it is physically meaningful, the entropy of
de Sitter, like any classical gravitational entropy, is a striking
example of non-decoupling, where a large-scale feature of a
gravitating system reflects microscopic details like the number of
degrees of freedom available to the theory.

In this note, we present a picture of Euclidean de Sitter space in
string theory which allows us to relate de Sitter entropy to the
number of degrees of freedom in a dual quantum mechanics.  This
leads to an argument that Euclidean de Sitter spaces in divers
dimensions may have dual descriptions in string theory in terms of
certain quantum mechanical models.

\section{Classic facts}

It is worth our while to begin with a review of known facts about
semiclassical de Sitter space\footnote{For example, see~\cite{he}.}.
De Sitter is a maximally symmetric solution to Einstein's equations
with a positive cosmological constant $\Lambda$.  Defining a length
scale $\ell = \sqrt{1/\Lambda}$, D-dimensional de Sitter space
($\ds{D}$) has a metric:
\begin{equation}
    ds^{2} = -dt^{2} + \ell^{2} \, \cosh^{2}(t/\ell) \,
    d\Omega_{D-1}^{2} \, .
\label{globalmet}
\end{equation}
Equal time sections of this metric, which covers $\ds{D}$ globally,
are (D-1)-spheres ($\sph{D-1}$).   These spatial sections have no
asymptotic region, and therefore there is no global notion of a
conserved energy in de Sitter space~\cite{mtw}.

Because of the exponential growth of the spatial sections at late
times, an inertial observer in $\ds{D}$ sees a cosmological
horizon.  The region of spacetime visible to such an observer can be
described by a static metric:
\begin{equation}
    ds^{2} = - V(r) \, dt^{2} + {1\over V(r)} \, dr^{2} + r^{2} \,
    d\Omega_{D-2}^{2}
    ~~~~~~;~~~~~~ V(r) = 1 - {r^{2} \over \ell^{2}} \, .
\label{staticpatch}
\end{equation}
Here polar coordinates have been constructed around an inertial
observer placed at $r=0$, which could be the north pole of the sphere
in (\ref{globalmet}).  The static patch displays an event horizon at
$r = \ell$, and the inertial observer has access to phenomena within
the region $r\leq \ell$. Essentially, de Sitter space can be viewed as
a finite cavity surrounding the observer, with the horizon as its
boundary.  While the metric (\ref{globalmet}) does not have any
globally defined timelike Killing vectors, the SO(D,1) isometry group
does give rise to a timelike Killing vector inside the static patch
(\ref{staticpatch}).  Abbott and Deser have shown that there is a
perturbative positive energy theorem with respect to this Killing
vector for excitations with support restricted to the static
patch~\cite{abbottdeser}.  Therefore, the static patch of de Sitter
space is perturbatively stable with respect to fluctuations that are
accessible to its inertial observer.

Applying the Bekenstein-Hawking entropy formula to the cosmological
event horizon yields an entropy
\begin{equation}
    S = {A \over 4 \, \gn} = {(2\ell)^{D-2}\, \pi^{(D-1)/2}\over \gn\,
   \Gamma((D-1)/2)} \sim
          {\Lambda^{-(D-2)/2}\over \gn} \, .
\label{sbh}
\end{equation}
Gibbons and Hawking showed that the classical laws of horizon
mechanics appearing in the physics of black holes apply equally to the
de Sitter horizon, so that {\it every} inertial observer sees a world
with horizons obeying the formal rules of thermodynamics~\cite{gh1}
with area playing the role of entropy.  It has been further argued
that the entropy of empty de Sitter space bounds the total
thermodynamic entropy of matter systems that can be placed within a
horizon volume of any world with a positive cosmological constant
$\Lambda$~\cite{raphael1,raphael2,raphael3}.

The analogy of static patch physics with thermodynamics is greatly
strengthened by the remarkable observation that detectors held by any
inertial observer in de Sitter space~\cite{gh1,nappi}, for example,
one stationed at $r=0$ in the static patch (\ref{staticpatch})
register a thermal bath with a temperature
\begin{equation}
    T = 
{1 \over 2\,\pi\, \ell}
    \, .
\label{dstemp}
\end{equation}
Every inertial observer in de Sitter space detects this temperature
and a horizon with entropy (\ref{sbh}) suggesting that these are
properties of the spacetime, and not of the frames occupied by
individual observers.  To clarify the situation it is helpful to
examine the thermodynamics of de Sitter space in semiclassical
Euclidean gravity.

\subsection{Euclidean de Sitter space and thermodynamics}

The Euclidean continuation of the static patch (\ref{staticpatch}) is
obtained by rotating $t \rightarrow it$ and identifying $t \sim t +
2\pi\ell$, in order to obtain a non-singular space.  Making the
coordinate transformations $t = \ell \, \tau$ and $r = \ell \,
\sin\theta$ gives the metric
\begin{equation}
    ds^{2} = \ell^{2} \, [d\theta^{2} + \sin^{2}\theta \,
    d\Omega_{D-2}^{2} + \cos^{2}\theta \, d\tau^{2}] = \ell^{2} \,
    d\Omega_{D}^{2}
\label{edsmet}
\end{equation}
In other words, Euclidean de Sitter (EdS) is a D-sphere with a round
metric.  The Euclidean horizon is the set of fixed points of the time
translation and is therefore a $(D-2)$-sphere of maximum size within
EdS.  The periodicity of Euclidean time implies a temperature
\begin{equation}
T = {1 \over 2\pi \ell} \, .
\end{equation}
The Euclidean continuation of global de Sitter time
($t \rightarrow it$, $t \sim t + 2\pi \ell$, $t =\ell \, \tau$)
similarly yields
\begin{equation}
    ds^{2} = \ell^{2} \, [d\tau^{2} + \cos^{2}\tau \,
    d\Omega^{2}_{D-1}] \, ,
    \label{edsmet2}
\end{equation}
which is the same as (\ref{edsmet}) in different coordinates;  thus the
round D-sphere is the unique Euclidean continuation of Lorentzian de
Sitter space.\footnote{Recall that global anti-de Sitter space (AdS)
and the Poincar\'{e} patch both continue to the same Euclidean
manifold. }

The Euclidean action with a positive cosmological constant is
\begin{equation}
    I = {1 \over 16\,\pi\gn} \, \int \sqrt{g} (R - 2\Lambda) \, .
\end{equation}
Since EdS has no boundary there are no boundary terms in the Euclidean
action.  As discussed by Gibbons and Hawking~\cite{gh2}, the Euclidean
approach to gravitational thermodynamics requires us to evaluate the
partition function with periodic boundary conditions in time:
\begin{equation}
    Z = \tr e^{-\beta H} = \int {\CD}\phi e^{-I(\phi)} \approx
    e^{-I_{cl}(\phi)} \, ,
\end{equation}
where $\phi$ includes all fields in the system and the last equality
represents the semiclassical limit.   In this limit, then,
\begin{equation}
    \ln Z \approx I_{cl} \equiv \beta(E - T\, S) \, .
\end{equation}
We can evaluate this on empty de Sitter space, using $R = 4\Lambda$
and the fact that $E = 0$ for a closed space like EdS (see,
e.g.,~\cite{mtw}).  This gives
\begin{equation}
    S \sim {(2\ell)^{D-2} \, \pi^{(D-1)/2}\over \gn \, \Gamma((D-1)/2)}
    \sim {\Lambda^{-(D-2)/2} \over \gn}
    \, .
\end{equation}
Since  both global de Sitter and the static patch are obtained as
continuations of the same Euclidean space,  this suggests that
de Sitter space is associated with an entropy $S \sim  \ell^{D-2}/\gn$.

\subsection{The meaning of entropy}
The purpose of this note is to analyze whether and how the entropy
of de Sitter space can be understood from string theory.   But what
does entropy mean in quantum gravity?     Recapitulating controversies over
black hole entropy, several interpretations are possible:
\begin{enumerate}
    \item dS entropy is just a formal analogy and should not be
    interpreted a real thermodynamic quantity.   We will set aside
    this skeptical position.\label{ent1}
    \item dS entropy arises by quantizing degrees of freedom
    associated with a horizon.   In $\ds{2+1}$, Carlip's horizon
    boundary conditions have yielded an account of de Sitter
    entropy~\cite{carlip,juanandy1}.\footnote{A different
    approach has appeared recently in~\cite{anastasia}.}     However,
    it is unclear at present how to extend this approach, relying on
    features of $2+1$ gravity, directly to higher dimensions.    For 
    some ideas in this direction see~\cite{hdimds}.
    \item dS entropy arises from quantum entanglement with
     degrees of freedom that are hidden behind the cosmological
    horizon.   If Newton's constant is wholly induced from matter
    fluctuations, black hole entropy can be understood in terms of
    entanglement~\cite{ent}.   In a brane-world scenario~\cite{ent2}
    this has been used to analyze de Sitter entropy.\label{ent2}
    \item dS entropy counts the number of initial conditions that can
    evolve into empty de Sitter space.\label{ent3}
    \item dS entropy counts the number of microscopic configurations
    that are macroscopically de Sitter.   This is the approach that
    succeeded for a large class of extremal and near-extremal black
    holes in string theory~\cite{stromvaf}.\label{ent4}
    \item dS entropy is the {\it finite} dimension of the Hilbert
    space describing the quantum gravity of de Sitter.  This radical
    conclusion has been particularly advocated by
    Banks~\cite{banks1}.\label{ent5}
    \item dS entropy counts the number of degrees of freedom of
    quantum gravity in de Sitter space.\label{ent6}
\end{enumerate}
Several of these philosophies have their roots in the principle of
holography proposed by 't Hooft and Susskind~\cite{holog}.

Note that there is a difference between (\ref{ent5}) and (\ref{ent6}),
since a single degree of freedom like a harmonic oscillator can have
an infinite dimensional Hilbert space.  Of course, at a fixed finite
temperature, only a finite dimensional Hilbert space will be
effectively accessible to a system with a finite number of degrees of
freedom.

In any case, holographic
arguments~\cite{holog,raphael3,raphael2,banks1,wittends,bf} strongly suggest
that de Sitter entropy is at least meaningful as a bound on the
functionally accessible entropy of matter systems within one horizon
volume of a world with a cosmological constant.  We might suppose that
this quantity is relevant to an inertial observer's effective
description of the world within a static patch, and that this is the
right way to think about de Sitter physics since there is a positive
energy theorem for the relevant fluctuations~\cite{abbottdeser}.  To
make sense of physics in global coordinates, it has been argued that
different static patches should be considered gauge copies of each
other so that dS entropy measures a property of the entire
universe~\cite{banks1,bf}.  In our approach, we will evade this
interpretational issue by simply discussing the entropy of de~Sitter
space in the {\it Euclidean\/} framework.

\section{Seeking de Sitter space}

\paragraph{No Go:}
Any attempt to realize de Sitter space within string theory must
contend with several no go theorems in the literature.  These include
one stating that de Sitter compactifications of conventional
supergravity cannot be found under the following
conditions~\cite{nogo2}: (a) The action does not contain higher
derivative terms, (b) The scalars have a potential $V(\phi) \leq 0$,
(c) The form fields have positive kinetic terms, (d) The Newton
constant on de Sitter is finite, namely gravitons interact with a
finite strength.  The low energy gravities arising from string theory
certainly have higher derivative corrections. Likewise, not all scalar
potentials arising in string theory are strictly non-negative.
However, we will not explore these avenues in this paper.  There is
also a theorem stating that 11d supergravity does not admit a
cosmological constant~\cite{nogo1}, and therefore does not give rise
to the 11d de Sitter space.  Finally, attempts to construct a maximal
($\CN = 8$) cosmological supergravity theory in four dimensions by
starting with an Einstein term and a positive cosmological
constant~\cite{nogo3} revealed that de Sitter superalgebras do exist,
but result in a Lagrangian with two closely related unpalatable
features: (a) the R-symmetry group $SO(6,2)$ is non-compact, and
therefore (b) the gauge field has some ghostlike components with the
wrong sign kinetic term.  Another curious feature of the de Sitter
superalgebra is that the supercharges square to zero~\cite{nogo3},
\begin{equation}
\sum \{Q,Q^\ast\}=0 \,,
\label{cureq}
\end{equation}
instead of anticommuting to the Hamiltonian, $\sum \{Q,Q^\ast\} = H$, as
usual.  Note that the absence of $H$ on the right hand side of
(\ref{cureq}) implies that a positive energy theorem cannot be proved
for de Sitter space by appealing to the supersymmetry algebra.  What
is more, the de Sitter superalgebras have no non-trivial
representation on a positive Hilbert space, which certainly precludes
any naive attempt to construct a Fock space of multi-particle states.
These facts seem to suggest a possible connection with topological
field theory.\footnote{A relation between de Sitter and topological
field theories has been explored by Hull~\cite{hull}.   The SO(6,2) de
Sitter gravity is also presented in these papers.}

\paragraph{Go?: } Hull and Warner showed that de Sitter spaces can
arise from compactification of supergravity on hyperbolic manifolds.
Problems with wrong sign kinetic terms were avoided in this approach
by squashing the hyperbolic space to remove non-compact
isometries~\cite{hullw}.  There have been attempts to escape these no go
theorems by changing the rul1es of the theories in which they were
derived.  Hull~\cite{hull} and Hull and Khuri~\cite{hullkhuri} have
discussed type II$^{*}$ strings and M$^{*}$ theory which can arise
from timelike T-dualities of the conventional systems.  These theories
can have a variety of different spacetime signatures and admit
Euclidean brane solutions whose near-horizon limits realize de Sitter
space.  In another approach, Chamblin and Lambert~\cite{andrewneil}
have discussed a massive form of the 11d supergravity equations that
admit de Sitter solutions.  It is unclear at present whether these
theories can be made stable and quantum mechanically consistent.

\paragraph{Go Around: }
We will evade the no go theorems reviewed above by not requiring that
dimensional reduction of the fundamental theory on a compact manifold
yields de Sitter space.  Rather, we are inspired by the result
of~\cite{nogo3} that a maximal supergravity on $\ds{4}$ would have an
SO(6,2) R-symmetry group.  This is suggestive because that SO(6,2) is
the isometry group of $\ads{7}$.  M-theory admits $\sph{4} \times
\ads{7}$ solutions in which the sphere is supported by a four-form
flux passing through it. In these compactifications, the isometry
group of the $\sph{4}$ acts as the SO(5) R-symmetry group of
supergravity on $\ads{7}$.  Turning things around we might expect that
a theory on $\sph{4}$ obtained by ``compactification'' of M-theory on
$\ads{7}$ should have an SO(6,2) R-symmetry arising from the AdS
isometry group.  Now recall that $\sph{4}$ is exactly the same
manifold as 4d, Euclidean de Sitter space, $\eds{4}$, i.e., $\sph{4}
\times \ads{7} \equiv \eds{4} \times \ads{7}$.  We therefore suggest
that:
\begin{center}
\begin{minipage}[c]{0.85\textwidth}
{\it
The $\sph{D} \times \ads{k+1}$ solutions of M-theory should be
regarded as $\eds{D} \times \ads{k+1}$ compactifications.
In this
picture, the cosmological constant is implemented by a D-form flux
passing through the de Sitter world.
}
\end{minipage}
\end{center}
We will examine this proposal in the four basic $\eds{D} \times
\ads{k+1}$ compactifications of M-theory, which have $(D,k+1) =
(5,5),(3,3),(4,7),(7,4)$.

At first sight there are two problems with this proposal.  First of
all, in usual compactifications, the Newton constant on the noncompact
factor is given by the higher dimensional $\gn$ divided by the volume
of the internal space.  AdS has an infinite volume and this threatens
to give an effective vanishing Newton constant on the EdS side.
Secondly, Lorentzian time is in the wrong place -- it is in the AdS
factor.\footnote{Possible compactifications on a Euclidean hyperbolic
space to a Lorentz-signature de~Sitter have been discussed by Hull
\cite{hull}.  This tempting ``double Wick rotation'' of the AdS
backgrounds requires Ramond-Ramond fields with the wrong-sign kinetic
terms, and therefore belongs to the realm of M${}^\ast$-theory.}  We
will address these issues in turn.

\subsection{Localization on AdS}

We are interested in the physics of fluctuations of EdS and not of
AdS. Therefore we should isolate a sector of the dynamics of $\eds{D}
\times \ads{k+1}$ which describes fluctuations of the sphere.  To do
this we focus on supergravity modes that are minimally excited on the
anti-de Sitter factor.  The metric of AdS creates an effective
potential well that localizes states of finite energy, although the
space is non-compact.   This is in marked contrast to flat space.
This feature implies that the modes of interest to us effectively
explore a finite volume of AdS.

To illustrate these points, it is helpful to examine the scalar wave
equation on $\eds{D} \times \ads{k+1}$. Placing a mode in a spherical
harmonic on the EdS factor yields a massive wave equation $(\nabla^{2}
+ m^{2}) \phi = 0$, where $m^{2}$ is the eigenvalue of the Laplacian
on the sphere.  Working with a global AdS metric
\begin{equation}
ds^2 = L^2 [ -\sec^2\rho \, dt^2 + \sec^2\rho \,
d\rho^2 + \tan^2\rho \, d\Omega^2_{k-1} ] \, ,
\label{adsmet}
\end{equation}
the normalizable mode solutions to the wave equation on AdS are:
\begin{eqnarray}
    \phi_{+} &\propto&  e^{-i\omega t} \,  Y_{lp} \, (\cos
    \rho)^{2h_{+}} \,  (\sin   \rho)^{l} \, \times \nonumber \\
     &{} & {}_2F_1(h_+ + (l+\omega)/2, h_+ + (l-\omega)/2, l + k/2;
    \sin^2\rho) \, , \label{modes}
     \\
    2h_{+} &=&  {k \over 2} + {\sqrt{k^{2} + 4 m^{2} L^2} \over 2} \, ,\\
    \omega L&=& 2h_{+} + l + 2n \, .\label{spectra}
\end{eqnarray}
Here $Y_{lp}$ is a spherical harmonic on the AdS factor of the space,
${}_2F_1$ is a hypergeometric function~\cite{bklt} and $L$ is the AdS
radius.  (Recall that in the $\ads{}\times \eds{}$ compactifications
of string theory $L$ is equal to $\ell$, the EdS scale, up to factors
of two.)  In particular, ground states, with $n = l =0$, have the form:
\begin{equation}
    \phi^{+} \propto e^{-i \omega t} \, (\cos \rho)^{2h_{+}}
\label{ground}
\end{equation}
and are therefore localized within one AdS radius.  It is easy to
check that as $l$ increases, the modes (\ref{modes}) are pushed
rapidly outward by the angular momentum barrier, and represent ripples
of AdS, and not fluctuations of EdS.  On the other hand, as $n$
increases, the wavefunction remains substantially localized inside the
AdS scale.  Higher harmonics on EdS increase the effective mass
$m^{2}$ and hence $h_{+}$, localizing the $l=0$ modes still further.
Although we have illustrated these point by studying scalar fields on
AdS, similar behaviour occurs for higher spin fields.  This analysis
suggests that all the physics of EdS is already contained in modes
that can be localized within one AdS radius.

One might prefer an alternative point of view in which we simply
impose a cutoff at one AdS radius.  In the CFT dual to AdS space, such
a cutoff should correspond to a reduction to a matrix
theory~\cite{hologbound,lenny1}.  We will later consider the
consequences of enlarging the cutoff.  However, we cannot decrease it
without cutting off part of the wavefunctions of ground state modes on
AdS.

Among the modes that we have admitted are the giant gravitons that
expand inside EdS.  These states, with $l=n=0$, are subject to a bound
on $h_+$~\cite{giant}, and expand into co-dimension two spheres on
EdS.  Interestingly, the largest giant graviton fills out the
Euclidean de Sitter horizon.

\subsection{``Compactifying'' on AdS}

We have argued that a finite spatial volume of AdS is accessible
during interactions between the modes of interest for EdS physics.
This does not yet imply a finite Newton constant on EdS, as we are
still left with the non-compact time dimension in the AdS factor.  In
order to complete the effective compactification on AdS we will now
make time in (\ref{adsmet}) periodic.

Supergravity modes on $\ads{k+1} \times \eds{D}$ are known to have
quantized frequencies in units of the AdS scale.\footnote{For
references, see the review \cite{adsrev}.}  In essence, this happens
because, despite its non-compactness, the AdS geometry creates an
effective potential well for the supergravity fluctuations.
Therefore, at the level of supergravity, we can compactify time with a
period $\sim L$.

Having made time periodic, we can also make it Euclidean at the same
period.   There are no normalizable mode solutions in Euclidean
time, and the frequencies of non-normalizable modes are not
quantized.  The compactification of time projects out many of these
modes, and only a quantized set of frequencies survives.    Earlier we
argued that the physics of EdS is already contained within the
normalizable modes which are localized within one AdS radius.
The analogous statement in Euclidean signature is that the
non-normalizable modes of interest show pure scaling behaviour outside
an AdS radius.

\paragraph{Estimating $\gn$: }
We have argued that all the physics of EdS is already contained in 
the physics of excitations within one AdS radius $L$.  The interaction
strengths of such modes are governed by the overlaps of their
localized wavefunctions on the AdS side.  Since these wavefunctions
are effectively contained in finite spatial volume $\sim L^k$, their
overlap can be estimated by cutting off the AdS space 
at one AdS radius and including the volume factor $L^k$ in the effective
interaction strength of harmonic modes on the sphere.  Moreover, 
the compactification of the AdS time at the scale $L$ implies that the
effective volume of the AdS space-time accessible to the modes of 
interest is finite, and scales as $L^{k+1}$.  Therefore, the effective
Newton constant for the excitations of the $\eds{D}$ factor is finite, 
and scales as
\begin{equation}
G_{D}\sim \frac{\gn}{L^{k+1}} \, 
\sim
\frac{\gn}{\ell^{k+1}} \, .
\label{effnewt}
\end{equation}
Here, and in future equations, we will only keep track of the scaling
of quantities, and will not attempt to determine numerical factors.

\section{De Sitter entropy}

Using our estimate of $\gn$ we can examine the entropy of the EdS factors
in the $\eds{} \times \ads{}$ compactifications of string theory.

\paragraph{{\bf EdS}$_5$ $\times$ {\bf AdS}$_5$: }
The geometric area of the Euclidean horizon of $\eds{5}$ in units of
the effective Newton constant $G_5$ gives the entropy:
\begin{equation}
S_5 \sim {\ell^3 \over G_5} \, .
\end{equation}
We have argued that because the modes of interest for the study of EdS
physics are effectively localized on AdS,
\begin{equation}
G_5 \sim {G_{10} \over \ell^5}  \, .
\end{equation}
The $\eds{5} \times \ads{5}$ compactification of IIb string theory is
dual to the superconformal Yang-Mills theory of the $SU(N)$ gauge
group, defined on the 4d $\ads{5}$ boundary. The $\eds{}$ radius is
given in terms of the string length and $N$ as~\cite{juanads} 
\begin{equation}
\ell \sim l_s (g_s \, N)^{1/4} \, .
\label{rel5}
\end{equation}
In these variables the entropy is given by
\begin{equation}
S_5 \sim N^2 \, .
\label{entropy5}
\end{equation}
All factors of $l_s$ had to cancel for dimensional reasons, but it is
significant that the string coupling $g_s$ drops out. Since the matrices of
the CFT are $N \times N$, the quantum mechanics obtained by
dimensionally reducing it to a point will have
\begin{equation}
N_{{\rm d.o.f.}} \sim N^2
\label{dof5}
\end{equation}
degrees of freedom.  In short, the gravitational entropy of $\eds{5}$ 
(\ref{entropy5}) matches the number of degrees of freedom
(\ref{dof5}) in the matrix mechanics.   This is in accord with the
expectation that the region in the deep interior of AdS, which
we have argued contains the modes relevant for EdS physics,
has a dual description in a matrix model~\cite{lenny1,hologbound}.

\paragraph{{\bf EdS}$_7$ $\times$ {\bf AdS}$_4$: }
In this case, the gravitational entropy is $S_7 \sim \ell^5/G_7 $, in
terms of $\ell$, the radius of $\eds{7}$ and $G_7$, the effective 7d
Newton constant.  Following earlier reasoning, $G_7 \sim l_p^9 \,
\ell^{-4}$, where $l_p$ is the 11-dimensional Planck length.  The
$\eds{7} \times \ads{4}$ compactification of M-theory is dual to the
strong coupling limit of the $SU(N)$ SYM in $2+1$ dimensions.  This is
a 3d CFT with $SO(8)$ R-symmetry arising from $N$ coincident
M2-branes. According to the AdS/CFT dictionary $\ell \sim N^{1/6}
l_p$~\cite{juanads}.  Assembling these facts, the gravitational
entropy of $\eds{7}$ is $S_7 \sim N^{3/2}$.  Although the dual CFT is
strongly coupled, the Bekenstein-Hawking entropy of non-extremal
2-branes suggests that there are $O(N^{3/2})$ independent degrees of
freedom in it~\cite{klebanov}.  Presumably, dimensionally reducing to
a point gives a quantum mechanics with $O(N^{3/2})$ degrees of
freedom, which matches the scaling of the $\eds{7}$ entropy.

\paragraph{{\bf EdS}$_4$ $\times$ {\bf AdS}$_7$: }
The gravitational entropy of $\eds{4}$ is $S_4 \sim \ell^2/G_4$.  As
discussed, $G_4 \sim l_p^9\, \ell^{-7}$.  The $\eds{4} \times \ads{7}$
compactification of M-theory is dual to the strongly coupled $(2,0)$
CFT in 6 dimensions.  This is the worldvolume theory of $N$ coincident
M5-branes.  The AdS/CFT dictionary states that $\ell \sim N^{1/3} l_p
$.  These relations yield an $\eds{4}$ entropy $S_4 \sim N^3$.  The
Bekenstein-Hawking entropy of the non-extremal M5-brane system
suggests that there are $O(N^3)$ degrees of freedom in the $(2,0)$
theory~\cite{klebanov}.  We would expect that dimensionally reducing
to a point yields a quantum mechanics with $N^3$ degrees of freedom,
matching the $\eds{4}$ entropy.

\paragraph{{\bf EdS}$_3$ $\times$ {\bf AdS}$_3$: }
Finally, consider the $\eds{3} \times \ads{3} \times {\cal M}$
compactifications of IIB string theory, with ${\cal M} = K3$ ($T^4$).
The gravitational entropy is $S_3 \sim \ell/ G_3$, where $\ell$ is the
radius of $\eds{3}$ and $G_3 \sim g_s^2 \, l_s^4 \, \ell^{-3}$.
Furthermore, $\ell^2 \sim l_6^2 \, \sqrt{N}$ where $l_6$ is the 6d
Planck length resulting from the compactification on $K3$ ($T^4$).
Putting these facts together, the gravitional entropy is $S_5 \sim N$.
The dual theory is the $(4,4)$ symmetric product sigma model on the
target $(K3)^N/S_N$ ($(T^4)^N/S_N$). The central charge of this theory
is $c \sim N$, and so we expect that the quantum mechanical reduction
has $\sim N$ degrees of freedom.  Again, this coincides with the
$\eds{3}$ entropy.

\paragraph{} 
We have argued that in a variety of examples, the number of degrees of
freedom of a quantum mechnical model derived from the AdS/CFT
correspondence matches the entropy of Euclidean de Sitter space.

\paragraph{} This picture makes one wonder whether there is a precise
dual to Euclidean de Sitter space in terms of a matrix quantum
mechanics.  We have suggested that the physics of Euclidean de Sitter
can be understood in terms of the lowest modes of fields which are
localized within one AdS radius.  In the dual CFT, this sector is
described by a quantum mechanical theory in which the spatial dynamics
has been completely suppressed.  There are two ways of
seeing this.  First of all, the duals to the bulk $l=0$ states 
are created by the lowest Fourier modes of CFT operators.  What is
more, the UV/IR connection in the AdS/CFT correspondence, and the
holographic renormalization group \cite{holorg,holorg1}, teach us that
phenomena localized deep in the AdS interior should be described
by the long wavelength limit of the CFT.  In particular, we expect
that the physics of EdS that we are trying to isolate is contained
within the quantum mechanical theory obtained by dimensionally
reducing the dual CFT to a point~\cite{lenny1}.

One can effectively view our calculation in this section as cutting
off AdS space at a radius $r \sim \ell$ to get a finite Newton
constant on EdS.  Instead of keeping the physics inside one AdS
radius, we could have chosen a different cutoff $\varepsilon \ell$,
effectively localizing inside a bigger portion of AdS.\/\  The
holographic argument of Susskind and Witten~\cite{hologbound} implies
that the number of CFT degrees of freedom required to describe this
region is then rescaled by $\varepsilon^{k}$.  This change of the cutoff
also rescales the effective $G_D$ on EdS by $1/\varepsilon^k$, and
therefore the Bekenstein-Hawking entropy is multiplied by
$\varepsilon^{k}$.  Therefore, the relation between the entropy of EdS and
the number of degrees of freedom in the quantum mechanical dual scales
correctly with changes in the cutoff.

\section{Discussion}

In summary, we have argued that Euclidean de Sitter space makes an
appearance in string theory as the sphere in the $\ads{} \times S$
compactifications.  The dynamics of EdS is carried by modes that are
localized on AdS, yielding a finite effective Newton constant on the
EdS factor. The resulting de Sitter entropy, measured as the area of
the horizon in units of this effective $\gn$, scales as the number of
degrees of freedom in the corresponding dual.

Our construction does not provide a decoupled theory of de Sitter
gravity.  Rather, the ambient AdS space acts like a regulator for the
de Sitter theory, in much the same way that the massive states of
strings regulate a theory of gravity that would otherwise be
ill-defined.   

The effective cutoff in AdS space certainly breaks conformal
invariance.  Since de Sitter space has a finite temperature, we also
expect that supersymmetry should also be broken by the choice of the
spin structure around the AdS time.  An alternative way of making the
Newton constant on EdS finite would be to orbifold AdS by a
sufficiently large discrete symmetry group to get a space of finite
volume.  Precedents in this regard include~\cite{garydon}.

It remains a challenge to continue our picture to Lorentzian de Sitter
space.  Since we started with Lorentzian time in the AdS factor, we
must either move it to de Sitter or deal with a theory with two times.
The second option arises if we rotate the EdS time back to Lorentzian
signature while leaving AdS untouched.  These geometries appear as
formal solutions of M${}^*$ and Type II${}^*$
theories~\cite{hull,hullkhuri}.  If we eliminate the Lorentzian AdS
time by Wick rotation, the Ramond-Ramond fields pick up wrong-sign
kinetic terms~\cite{hull,hullkhuri}.   Clearly, there are questions
about the stability and the quantum mechanical definition of these
theories.   Still, one is left with a sense that de Sitter is asking
for two times.  (In this regard, recall the non-compact R-symmetry
groups in de Sitter superalgebras~\cite{nogo3}.)

An alternative approach is to attempt to make sense of the one
Lorentzian time that is already present on the AdS side, perhaps by
``grafting'' it onto the EdS time.  Consider choosing a new generator
of Lorentzian time translations $H' = H + J$, where $H$ is the
Hamiltonian on AdS and $J$ is the EdS rotation generator corresponding
to Euclidean time translations.  This procedure is somewhat
reminiscent of topological twisting, since it mixes the original
Hamiltonian with an R-symmetry generator.

If we can make sense of our picture in Lorentzian signature, it would
be natural to attempt an interpretation of de Sitter entropy in terms
of a counting of states.  Which states should be counted?  There is a
natural bound on the energy of excitations that can be localized
within a de Sitter volume. This bound is given by the mass of the
largest black hole that can be placed within a horizon volume.  The
mass of the D-dimensional extremal Schwarschild-de Sitter solution,
known as the Nariai black hole~\cite{nariai}, is given by
\begin{equation}
M \sim {\ell^{D-3} \over G_D} 
\end{equation}
In the CFT variables of our scenario $\ell^{D-2} \over G_D$ scales
precisely as the number of degrees of freedom, or central charge $c$,
of the dual theory.  Therefore, the Nariai bound for the bulk energy
translates into a bound on dimensionless CFT energies of the form $E
\equiv \ell M \leq c$.  This picture would suggest that the entropy of
de Sitter space should be interpreted in terms of the finite dimension
of the Hilbert space of states satisfying the Nariai bound.

\vspace{0.25in}
{\leftline {\bf Acknowledgements}}

We have benefitted from conversations with many colleagues including
Tom Banks, Jan de Boer, Raphael Bousso, Sean Carroll, Robbert
Dijkgraaf, Willy Fischler, Rajesh Gopakumar, Aki Hashimoto, Chris
Hull, Sunny Itzhaki, Antal Jevicki, Igor Klebanov, Amanda Peet, Arvind Rajaraman,
Lenny Susskind, Herman Verlinde, and Edward Witten.  We thank Chris
Hull for useful comments on this manuscript.  {\small V.B.}, {\small
P.H.}, and {\small D.M.} are supported respectively by the DOE grants
DE-FG02-95ER40893, DE-FG02-96ER40959, and DE-FG03-84ER40168.  In final
stages of this work we enjoyed the hospitality of the ITP, Santa
Barbara, and were supported by NSF grant PHY99-07949.  The results in
this paper were presented by one of us ({\small P.H.})  at ``Strings
2001'', in Mumbai, India.  We are grateful to the organizers for a
stimulating conference.

\vfill


\end{document}